# Improved Task Scheduling for Virtual Machines in the Cloud based on the Gravitational Search Algorithm

Basilis Mamalis
University of West Attica
Ag. Spyridonos, 12243, Athens, Greece
vmamalis@uniwa.gr

Marios Perlitis
Democritus University of Thrace,
Campus, 69100, Komotini, Greece
mperlitis@gmail.com

## ABSTRACT
The rapid and convenient provision of the available computing resources is a crucial requirement in modern cloud computing environments. However, if only the execution time is taken into account when the resources are scheduled, it could lead to imbalanced workloads as well as to significant under-utilisation of the involved Virtual Machines (VMs). In the present work a novel task scheduling scheme is introduced, which is based on the proper adaptation of a modern and quite effective evolutionary optimization method, the Gravitational Search Algorithm (GSA). The proposed scheme aims at optimizing the entire scheduling procedure, in terms of both the tasks execution time and the system (VMs) resource utilisation. Moreover, the fitness function was properly selected considering both the above factors in an appropriately weighted function in order to obtain better results for large inputs. Sufficient simulation experiments show the efficiency of the proposed scheme, as well as its excellence over related approaches of the bibliography, with similar objectives.

## 1. INTRODUCTION
During the last decade, cloud computing has evolved into an extremely popular (used both publicly and commercially) computing paradigm. In this context, a cloud computing platform can provide efficiently plenty of services to users with VM infrastructure as the main component [1,2]. For the end users all underlying resources are transparent. Actually, each submitted job occupies an independent VM. The relevant computing and memory resources are properly set in a mutually isolated status [3]. In such environments, each physical machine can host many VMs, thus ensuring that all users' applications will run independently and efficiently. As a result, in cloud computing platforms, task scheduling is actually taking place among the available VMs. Moreover, preserving load balancing among the VMs is still a difficult problem in such infrastructures.

In the above context, one of the most significant problems in modern cloud computing environments is task allocation. The main advantage of modern cloud platforms is the ability to be sufficiently elastic, i.e. being rapidly grown and teared down at any moment. This goal needs the support of novel task allocation algorithms to optimize key parameters like the total makespan, the responses times and the average utilisation of the resources. The function of scheduling in modern cloud computing is to allocate tasks on VMs in such a way that makespan and response times are minimized and resource utilisation is maximized.

Obviously, finding the proper mappings between the jobs submitted by end users and the dynamic resources of the available VMs is an NP-hard optimization problem. For this problem, a variety of dynamic, static and mixed scheduling algorithms have been proposed [4-10]. Some of the well-known early static scheduling schemes were based on ISH, MCP and ETF algorithms [8-10]. Most of these algorithms were built for the BNP class and were suitable for high efficiency network platforms in distributed environments. However, the relevant needs in cloud infrastructures are more advanced, whereas the costs of the provided services usually depend on the amount of usages. As a consequence, the typical early static algorithms can't play significant role.

Provided that the needs for efficient job/task assignments in cloud platforms have rapidly evolved, it is crucial to strengthen the total efficiency of the underlying system infrastructure. Combining system efficiency and optimized tasks execution time is a hard problem for task scheduling in cloud platforms. Some of the more recent improved scheduling algorithms used for addressing the above problems were SJF, FCFS, MET, MCT, Min-Min, Max-Min, LBSM etc. [11-27]. Recently, the swarm intelligence algorithms (like PSO) have also been used effectively towards the above direction [28-30]. A more comprehensive analysis of the most important of the above schemes is given in the following section.

In the present work, a new job/task scheduling scheme is introduced using the Gravitational Search Algorithm (GSA), i.e. a quite recent and effective evolutionary optimization method. The basic purpose of the given scheme is to optimize the whole scheduling procedure, in terms of both the total completion time of the scheduled tasks and the system resource utilisation. One of the key issues of the proposed method is the proper adaptation of the relevant fitness function, which is chosen considering both the above crucial factors (execution time, resource utilisation) in a properly weighted manner, in order to get better results for large inputs. The corresponding simulation experiments show the effectiveness of the proposed approach, and its excellence over other related approaches in the bibliography (like OLB [17], Min-Min [19], ESTA [27] and improved PSO [30]).

The remainder of the paper has been organized in the following form: In the next section (Section 2), the most recent related work is presented. Section 3 gives the necessary knowledge with respect to the GSA method. In section 4, the proposed

GSA-based task scheduling scheme is further presented and properly documented. Section 5 demonstrates the experimental evaluation of the proposed scheme, whereas section 6 concludes the paper.

## 2. RELATED WORK

As referred previously, job/task scheduling schemes are mainly distinguished in *static* and *dynamic*. With respect to static scheduling, all the necessary info for tasks execution (such as task execution time, VM capacity, weight of task etc.) is known before execution, whereas in dynamic scheduling, the decision is taken during runtime, according to probably changed parameters during the execution [11]. Both the above task scheduling types can be used on homogeneous and heterogeneous systems.

The task scheduling algorithms can also be distinguished into dependent and independent. In independent job/task scheduling no dependencies are assumed among the tasks [21-25]. All the tasks may be executed separately, and there is no need to keep any precedence log among them. On the contrary, considering dependent job/task scheduling specific precedence log must be maintained and relevant constraints must be satisfied before allocating the tasks to VMs [26].

Trying to address the most known independent job/task scheduling schemes of the literature, one should definitely refer to classical approaches like OLB, MET [17-18], Min-Min, Max-Min [19-20], LBSM [21]), MCT [22], Sufferage [23], LBMM [24], LJFR-SJFR [25], and RDLBS [26], as well as to more recent ones (based on modern combined techniques) like ESTA [27] and improved PSO [30].

In more details, OLB [17] allocates each task at random to subsequent nodes/VMs, provided they are available. The main objective of OLB is to achieve the maximum utilisation of the available VMs by keeping busy all the nodes involved in the system. However, it usually leads to poor makespan, since it allocates the tasks to particular nodes without taking in account their expected execution time. On the contrary, the main objective of MET algorithm [18] is to place the tasks with minimum expected execution time to the available VMs without considering the availability of the nodes. Since also the assignment of the tasks is mainly targeted to the faster nodes, the whole scheduling scheme offers higher performance. However, due to the above fact, the entire allocation leads also to significant imbalance of the total load, especially in the case of heterogeneous systems. Similarly, the main objective of MCT [22] is to allocate the task with the minimum expected completion time to the available VMs and along with it also allocate the tasks that don't have minimum completion time for the given resources. Since a portion of the tasks is to be assigned to resources without having minimum completion time, MCT combines the main features of both OLB and MET, and it naturally leads to better performance.

Considering the Min-Min heuristic [19-20], first, all the necessary info related to each task is supposed to be known at first. The Min-Min procedure is executed in two stages as below: (a) First, the MCT for each task is found, next the minimum time to completion of all tasks present in the batch is calculated; (b) Finally, the task with the minimum time to completion is chosen and then it's allocated to the proper node and removed from the batch when it completes. The Min-Min procedure is one of the easier methods to efficiently assign tasks to the available resources. However, it also has a significant disadvantage; specifically, the largest in size tasks will probably be waiting for undesirably long time. Further, the LBMM [24] heuristic stands as the combination of Min-Min and OLB scheduling schemes, thus leading to even better performance on makespan.

On the other hand, the Max-Min algorithm [19-20] also works into two stages, with a small difference: (a) First, the MCT for each task is found, then the maximum time to completion of all tasks present in the batch is calculated; (b) Finally, the task with the maximum time to completion is chosen and then it's given to the proper node and removed from the batch. The Max-Min scheme is usually the basis for such algorithms in distributed environments. The Max-Min approach represents the method of mapping tasks on the faster VM/nodes, having the tasks with longer completion times to run first. So, it is naturally expected to behave better than the Min-Min approach.

In Sufferage [23], the work is also separated in two stages. Firstly, the MCT is computed and then the second MCT is also computed for each task. Finally, the difference between the first and the second MCT values is also computed. This difference is the sufferage value; next, the task with the largest sufferage value is determined and then it's assigned to the appropriate resource with MCT.

Considering the LJFR-SJFR algorithm [25], first, a group of not-mapped tasks is initiated and also the relevant MCTs are determined. Next, the task with global MCT is chosen from the set of MCTs, and it's considered as SJFR. Also, the task with the global maximum completion time value of the tasks is determined, and it's considered as LJFR. Finally, shorter and longer tasks are assigned to the nodes.

Moreover, in LBSM [21] the authors present an efficient strategy for load balancing, and then the whole algorithm executes on specific inter-connection network. Initially, the assignment of the task is performed at random. After the initial task allocation, migration of specific tasks may take place from over-loaded nodes to under-loaded ones, to preserve load balancing. The above two strategies are almost similar, with the migration strategy being their only difference. LBSM performs better in terms of load balancing and computing components utilisation.

In [27], a new scheme (ESTA) is introduced to schedule tasks with no dependences in VMs with heterogeneous resources. The main objective of ESTA is to push the utilisation of all the VMs to the maximum (acting as an IaaS cloud computing provider), and also push the makespan to the minimum (as well as the average response time of the tasks). The incoming tasks are divided by the ESTA scheduler into largest size and smallest size tasks. The VM manager provides the set of VMs allocated to physical machines. ESTA is experimentally evaluated and compared to Min-Min and other competitive approaches of the bibliography to show its important performance gains.

Finally, in [30] an improved heuristic approach based on the PSO algorithm is presented to solve the problem of load balancing in VMs. By establishing relevant knowledge between the tasks and the VMs effectively, the main objective is the calculation of an optimal scheduling pattern, that not only minimizes the makespan, but also maximizes the VMs utilisation. Also, a new inertia weight technique is adopted through the classification of the fitness values. The corresponding simulations indicate the fast convergence speed of the algorithm, as well as its high efficiency and its great practical effect. Generally, the use of advanced heuristic algorithms with carefully selected fitness functions seems to be quite promising and effective even when the problem continues to expand, preserving scalability.

# 3. THE GRAVITATIONAL SEARCH ALGORITHM

GSA is a relatively new optimization algorithm based on the law of gravity and mass interactions. In the corresponding algorithm, the searcher agents are a collection of masses which interact with each other based on the Newtonian gravity and the laws of motion. A detailed description of the GSA approach is given in [3]. Let's suppose that initially there is set of agents, $S_A$. Each one of these agents is expected to eventually offer a portion of the final solution. Let's also denote the location of agent $L_i$, $1 \leq i \leq S_A$ in dimension $k$ as $x_i^k$ and the velocity of agent $L_i$ as $v_i^k$, $1 \leq k \leq K$. It is also assumed that each agent has the same dimension. Each one of the agents is then evaluated to verify the eligibility of the result using a specific fitness function. Moreover, if the $i^{th}$ location of an agent $L_i$ is being represented as $X_i = (x_i^1, x_i^2, ..., x_i^K)$, then the following formula represents the force applied on the $i^{th}$ agent by the $j^{th}$ agent.

$$F_{ij}^k(t) = G(t)\frac{M_{pi}(t) \times M_{\alpha j}(t)}{R_{ij}(t)}(x_j^k(t) - x_i^k(t))$$

In the above expression, $M_{pi}$ denotes the passive mass of the $i^{th}$ agent, $M_{\alpha j}$ denotes the relevant active mass of the $j^{th}$ agent, whereas $\alpha$ is a constant. Also, $G(t)$ is defined to be equal to $G_0(t_0/t_{max})^\beta$, in which $G_0$ stands also as a constant. $R_{ij}(t)$ denotes the Euclidean distance from $i^{th}$ agent by $j^{th}$ agent. So, the total force applied by the set of agents over agent $i$ in dimension $d$ is defined as given below.

$$F_i^k(t) = \sum_{j=1, j \neq i}^{S_A} rand_j \times F_{ij}^k(t)$$

The gravitational and the inertial mass are calculated by evaluating the fitness function. The greater the mass of an agent the greater the efficiency of the solution it represents.

$$m_i(t) = \frac{fit_i(t) - worst(t)}{best(t) - worst(t)} + \varepsilon$$

Here, $\varepsilon$ represents a limited constant, $fit_i(t)$ is the fitness of the $i^{th}$ agent, and $worst(t) / best(t)$ are as follows.

$$best(t) = \min_{j \in \{1...S_A\}} fit_j(t) \quad worst(t) = \max_{j \in \{1...S_A\}} fit_j(t)$$

$$M_i(t) = \frac{m_i(t)}{\sum_{j=1}^{S_A} m_j(t)}$$

In the above, it's also assumed that $M_{pi}$, $M_{ai}$, $M_i$ and $M_{ii}$ are all equal to each other. According to the $2^{nd}$ law of Newton, the following formula holds.

$$a_i^k(t) = \frac{F_i^k(t)}{M_{ii}(t)}$$

More concretely, the inertial mass of the $i^{th}$ agent is given by $M_{ii}$, and it's acceleration is given by $a_i^k(t)$.

# 4. THE PROPOSED SCHEDULING SCHEME

As it was also discussed, in cloud platforms the user-submitted jobs to the available VMs can be distinguished in two basic types: *independent* and *interrelated*. Moreover, the interrelated jobs can usually be spread into multiple small independent tasks that may run separately (without communicating with each other), so it is sufficient to address how the proper balance of the VMs workload is preserved with independent tasks only.

The main objectives of the proposed here solution are to achieve the maximum resource utilisation of the VMs along with the minimum execution time of the tasks. It has also been assume (as in [30]) that the VM is the basic resource unit and the demands of each task for resources is the same for all VMs. This assumption is somewhat against the best behavior of some of the competitive approaches (like ESTA and Min-Min), however it's a good starting point for comparison.

More concretely, it's assumed that there are $m$ VMs inter-connected adequately. The set $V$ of the available VMs can be represented as $(v_1, v_2, ... v_m)$, where $v_i$ denotes the max amount of resources of VM $i$, with $i=1...m$. Also, there is a set $T$ of tasks which can be represented as $(t_1, t_2, ... t_n)$, where $t_j$ denotes the task $j$, with $j=1...n$. Each task $t_j$ is further represented by a pair of values *(timeReq, resourceReq)*, in which *timeReq* denotes the required execution time of $t_j$, and *resourceReq* denotes the resource requirements of $t_j$.

The design of the proposed task scheduling scheme has been partially based on the way the GSA algorithm is utilised in [31], and it is suitably adapted over the total set of tasks to be scheduled. The fact that the set of the tasks is of controlled size makes possible the centralized use of fast evolutionary techniques that are highly accurate. Further, the proposed adapted GSA-based algorithm is executed to associate subsets of tasks to specific VMs for their eventual execution. Since also the maximization of the resource utilisation of the available VMs is of high priority, the fitness function was suitably modified / optimized to achieve proper balancing and get better results for large and very large inputs.

Specifically, the main set of agents ($S_A$) is first initialized; it must be noted here that the agents themselves stand as potential solutions. Moreover, as mentioned above, in the present context of scheduling the agents represent the associations of tasks to relevant VMs. So, assuming that $L_i$ is agent $i$, each item $x_i^k(t)$ assigns the corresponding task to a VM and $1 \leq i \leq S_A$, where $1 \leq k \leq K$ ($K$ equals to $n$). Thus, the following expression could represent agent $i$ [31], and the adapted GSA algorithm could then execute as follows (see next page, Algorithm 1)..

$$L_i = [x_i^1(t), x_i^2(t), x_i^3(t), ..., x_i^K(t)]$$

## 4.1 Determining the Fitness Function

The fitness function ($f$) has to be suitably determined taking in account two main factors:

(a) The total execution time of all the tasks in the available VMs (let's name it $T_{exec\_max}$), which equals to the maximum time required among all the VMs to execute its own tasks. So, if it's assumed that each *VM i* ($i=1...m$) has been assigned a set of $N_i$ tasks (denoted as $t_{ij}$, where $j=1...N_i$), $T_{exec\_max}$ can be expressed as follows:

$$T_{exec\_max} = max_{i=1}^{m}(\sum_{j=1}^{N_i} t_{ij}.timeReq)$$

(b) The total resource utilisation of the VMs during the execution of all the tasks (let's name it $R_{util\_sum}$), which equals to the sum of the resource utilisations of all the VMs, having each of them executing its own tasks. So, if it's assumed here again that each *VM i* ($i=1...m$) has been assigned a set of $N_i$ tasks (denoted as $t_{ij}$, where $j=1...N_i$), $R_{util\_sum}$ can be expressed as follows:

$$R_{util\_sum} = \sum_{i=1}^{m}(\frac{\sum_{j=1}^{N_i} t_{ij}.resourceReq}{v_i})$$

---

**Algorithm 1. GSA-based Task Scheduling**

---

**Input:**

Set of Tasks to be scheduled: $T = \{t_1, t_2, t_3, \ldots, t_n\}$

Set of VMs for tasks execution: $V = \{v_1, v_2, v_3, \ldots, v_m\}$

Set of initial agents, having size equal to $S_A$

Dimension of each agent = # of Tasks = $n$

**Output:**

The optimized tasks assignments to specific VMs

---

**GSA_scheduling_procedure:**

Agent $L_i$ is initialized, $\forall i$, $1 <= i <= S_A$

The assignment function is defined (for every $t_k$ to a $v_j$)

**do**   /* suppose at the beginning t=0 */

  **for** $i=1$ to $S_A$

    The fitness value $L_i$ is calculated

    The values *best* and *worst* are updated for all $L_i$

    $a_i^k(t)$ and $M_i(t)$ are computed for each agent $L_i$

    The position and the velocity of $L_i$ are updated

  **endfor**

**while** the termination criteria are not satisfied

---

**Fitness Function**

---

**Objective 1:**

$$\text{Minimize } T_{exec\_max} = max_{i=1}^{m}\left(\sum_{j=1}^{N_i} t_{ij}.timeReq\right)$$

---

**Objective 2:**

$$\text{Maximize } R_{util\_sum} = \sum_{i=1}^{m}\left(\frac{\sum_{j=1}^{N_i} t_{ij}.resourceReq}{v_i}\right)$$

---

**Total Objective:**

$$\text{Minimize } f = \frac{\gamma \times T_{exec\_max} + \varepsilon 1}{\delta \times R_{util\_sum} + \varepsilon 2}$$

---

As it was also discussed in the previous sections, in a modern cloud computing environment (datacenter) the minimization of the total execution time of the tasks is not the only priority. The maximization of the utilisation of the VMs involved is also an equivalently important performance factor. In many cases the tasks are accompanied with specific time limits that have to be satisfied; these limits represent also the highest expected task execution time by the cloud VM infrastructure (i.e. faster execution times make no sense). In such environments, the maximization of the VMs utilisation may reliably lead to significant savings of VM resources (preserving in any case the execution time of the tasks in the desirable levels), which then could mean that more tasks/jobs may be admitted and executed in time by the same/entire VM infrastructure. Furthermore, through the specific fitness function an optimized approach is proposed, aiming at the maximum possible balance on the actual behavior of the two main factors discussed above. Note that $\gamma,\delta$ and $\varepsilon_1,\varepsilon_2$ are suitably selected constants (dependent and independent to the total tasks execution time / resource utilization, respectively). Moreover, the function value is normalized appropriately within [0,1] to optimize the result. The fitness function proposed here, does not only properly distribute the weight of both the basic factors referred above (resource utilization and execution time) in the overall calculation. It also gives the developer the capability to fix/repair (using $\gamma,\delta$) adequately the non-unpredictable (as well as non-canonical) differences caused by the potentially different measure units. It also offers the convenience to incorporate (using $\varepsilon_1,\varepsilon_2$) in the optimization procedure other important factors / parameters too. Other types of fitness functions [31] could also be applied and evaluated, which is of high priority in future research.

## 5. PERFORMANCE EVALUATION

In the following, a set of extended simulation experiments is presented in details, performed to estimate the efficiency of the proposed GSA-based task scheduling algorithm in comparison to OLB [17], Min-Min [19], ESTA [27] and improved PSO [30]. The first two of the above algorithms (OLB, Min-Min) are typical representatives of early static task scheduling algorithms, whereas the rest (ESTA, PSO) are two of the more recent highly effective task scheduling algorithms based on modern techniques. The Cloudsim 4.0 simulator [31] has been used, deploying small to large scale random test inputs. Each VM has 1GHz CPU, 16-64GB memory and 1000 Mbits/s bandwidth. To check the performance of the GSA-based algorithm, experiments involving from 16 up to 1024 VMs and from 1000 up to 50000 tasks were performed.

### 5.1. Results for varying number of tasks

In the first set of experiments, the influence of varying values in the number of tasks to fixed number of VMs is examined on total performance. More concretely, it's assumed there are 16 VMs, and a varying number of tasks is deployed; specifically ranging from 1000 up to 50000 tasks. The main observations are as follows.

As the number of tasks increases, makespan on the VMs also increases, as shown in Fig. 1. Moreover, as seen in Fig. 1, the proposed GSA-based scheduling approach performs better on makespan among all the considered competitors. Specifically, the average performance gain of GSA over ESTA, PSO, Min-Min and OLB, is 5.1, 9.4, 6.8 and 31.3%, respectively. When there are more varying tasks at the same number of VMs, ESTA is very close to Min-Min, quite better than PSO and clearly better than OLB strategy; which has the worst performance on makespan, as it was expected due to the fact that it assigns the tasks to VMs without taking in account their expected execution time.

Furthermore, in Fig. 2 the results of average utilisation of the system are shown for each case. Here again, the first observation is that, as it was normally expected, the average utilisation progressively increases (up to 97.3% for 50000 tasks if the GSA algorithm is considered) as the number of tasks increases. Moreover, the performance of the proposed GSA-based scheduling algorithm remains the best among all the considered heuristic strategies. More concretely, the average utilisation improvement of GSA over ESTA, PSO, Min-Min and OLB, is 4.7, 7.2, 10.7 and 43.5%, respectively. Also, when there are more varying tasks at same number of VMs, ESTA is very close to PSO, quite better than Min-Min and substantially better than OLB strategy.

Note here that the improved PSO approach behaves better than Min-Min, since the maximization of the system resource utilisation is inherently one of its main objectives. On the contrary, the Min-Min algorithm behaves better on makespan; which has the main priority in its design objectives.

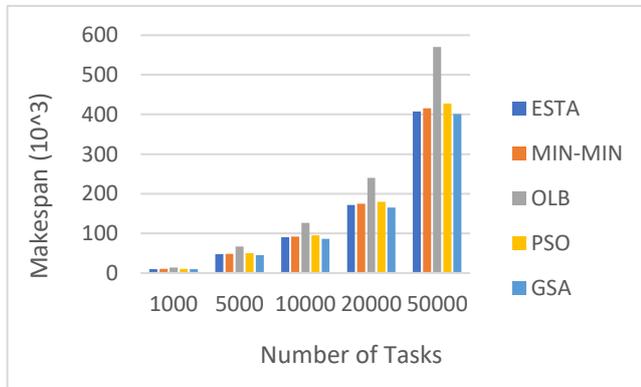

**Fig 1.** Makespan for varying # of tasks

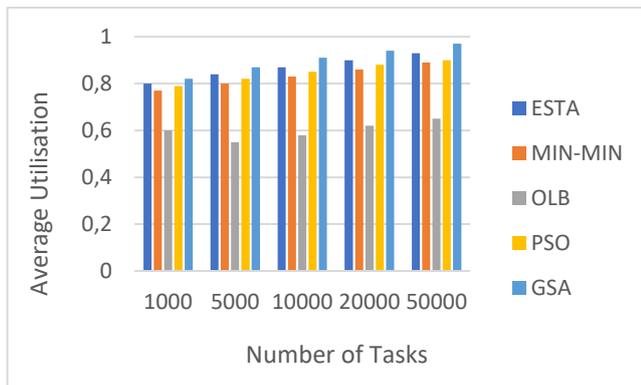

**Fig 2.** Average utilisation for varying # of tasks

## 5.2. Results for varying number of VMs

In the second set of experiments, the influence of different values of the number of VMs to fixed number of tasks, is examined on total performance. More concretely, it's assumed there is a fixed number of 50000 tasks, whereas the number of VMs (in which the tasks will be deployed) varies from 16 to 1024. The main observations are as follows.

As the number of VMs increases, both the makespan and the average utilisation values decrease, as shown in Fig. 3-4. Moreover, as indicated in Fig. 3, the proposed GSA-based scheduling approach performs better on makespan among all the considered heuristic strategies. Specifically, the average performance gain of GSA over ESTA, PSO, Min-Min and OLB, is 3.5, 7.7, 5.1 and 27.3%, respectively. Additionally, when there are more VMs at the same number of tasks, ESTA is very close to Min-Min, quite better than PSO and clearly better than OLB strategy; which has the worst performance on makespan for the same reasons as it was explained above (in the context of the first experiment).

Furthermore, in Fig. 4 the results of average utilisation of the system are shown for each case. Here again, the first observation is that, as it was normally expected, the average utilisation progressively decreases (down to approximately 90% for 1024 VMs if the GSA algorithm is considered) as the number of VMs increases. Moreover, the performance of the proposed GSA-based scheduling algorithm remains the best among all the considered heuristic strategies. More concretely, the average utilisation improvement of GSA over ESTA, PSO, Min-Min and OLB, is 6.7, 9.2, 14.5 and 68.7%, respectively. Having also more varying VMs at the same number of tasks, ESTA is very close to PSO, quite better than Min-Min and substantially better than OLB strategy. Note here also (as it was observed for varying number of tasks too in the first experiment) that the improved PSO approach behaves better than Min-Min, since the maximization of the system resource utilisation is inherently one of its main objectives.

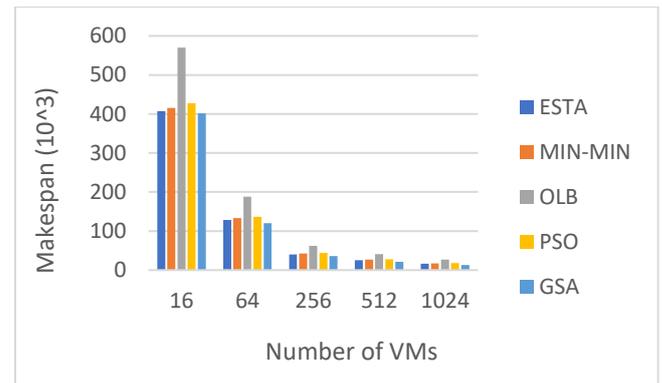

**Fig 3.** Makespan for varying # of VMs

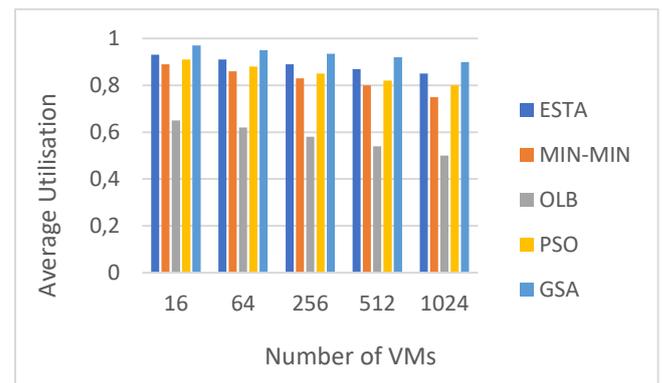

**Fig 4.** Average utilisation for varying # of VMs

## 5.3 Comparing to Improved PSO

As mentioned in the previous section, the comparison of the proposed GSA-based approach with the improved PSO approach [30] is the most fair and representative one, since the two algorithms not only have the same objectives, but also adopt exactly the same assumptions.

Moreover, they both come from the same base area of the evolutionary optimization algorithms, i.e. they may be regarded as two competitive approaches from the same 'family'. In the above context, beyond the results shown in Fig. 1-4, the corresponding makespan and utilisation curves for only the GSA-based and improved PSO approaches are presented also in Fig. 5-8, for direct comparison.

As it can be easily observed in Fig. 5, the makespan achieved by the GSA-based approach is clearly better than the makespan of improved PSO scheme, especially for large number of tasks (i.e. in cases of 20000 and 50000 tasks). For smaller numbers of tasks the improved PSO algorithm behaves quite efficiently and achieves makespan values that are very close to the ones of the GSA-based algorithm.

Moreover, in Fig. 6 the corresponding utilisation curves are presented. As shown in Fig. 6, as the number of tasks increases, the average resource utilization increases too for both approaches. Furthermore, the relevant increase for the GSA-based approach is sharper and finally leads to a much better utilization value (0.97) for 50000 tasks. Here also again, the improved PSO algorithm behaves more efficiently for smaller numbers of tasks, leading to utilisation values quite close to the ones of the GSA-based algorithm.

Additionally, in Fig. 7 the makespan curves are presented for varying number of VMs. As it can be easily observed here also, the makespan achieved by the GSA-based approach is clearly better than the makespan of the improved PSO scheme, especially for smaller number of VMs (i.e. in cases of 16 and 64 VMs). Actually, the percentage difference between the two approaches is approximately the same for all the varying numbers of VMs; and just the absolute values difference is more visible in the case of smaller number of VMs.

The above conclusion may also be drawn by observing Fig. 8, where the corresponding average utilisation curves are given. More concretely, as it can be seen in Fig. 8 the average resource utilization achieved by the GSA-based approach is clearly better than the one of the improved PSO approach, for all the numbers of available VMs. However, the improved PSO algorithm behaves quite efficiently too, and achieves very satisfactory utilisation values even for large number of VMs (0.85 for 1024 VMs vs. 0.91 achieved by GSA).

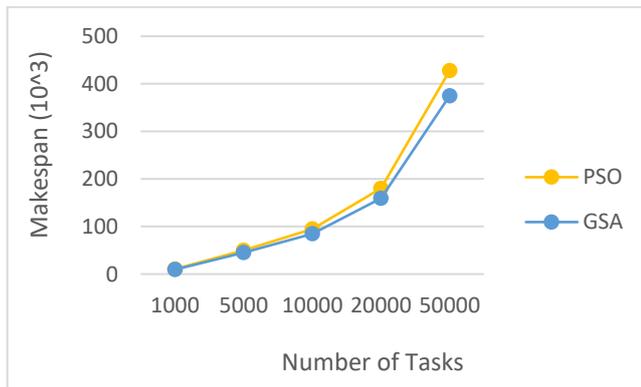

**Fig 5.** Makespan curves for varying # of tasks

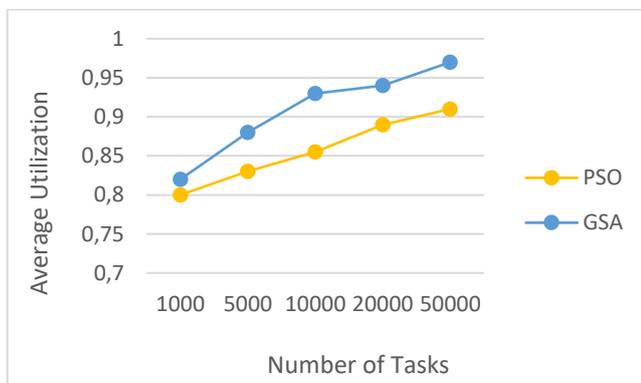

**Fig 6.** Utilisation curves for varying # of tasks

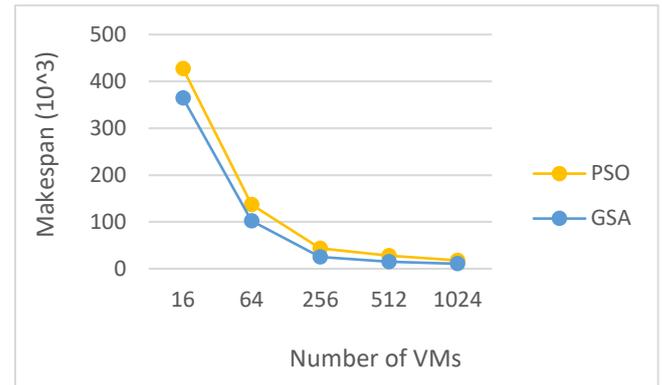

**Fig 7.** Makespan curves for varying # of VMs

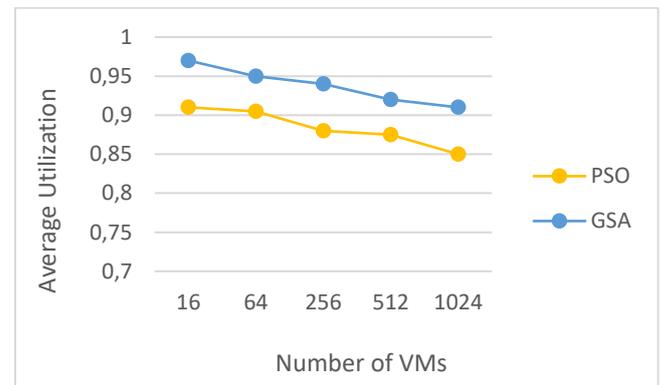

**Fig 8.** Utilisation curves for varying # of VMs

## 6. CONCLUSION

An efficient task scheduling scheme based on the Gravitational Search Algorithm (GSA) is presented and evaluated throughout the paper. The proposed method is a novel approach that exploits the effectiveness of a modern and quite efficient evolutionary optimization technique, to optimize the basic task scheduling procedure in cloud computing environments. It considers both the total tasks execution time and the system resource utilisation as the critical factors for optimization, and adapts adequately the relevant fitness function in order to obtain sufficiently accurate solutions even for large inputs. The experimental evaluation of the proposed scheme shows its high efficiency in terms of makespan and resource utilisation, as well as its superiority over other competitive approaches of the literature (like OLB, Min-Min, ESTA and improved PSO). The suitable adaptation and evaluation of alternative types of fitness functions is of high priority in future research.